\documentstyle[multicol,prd,aps,graphicx]{revtex}

\begin{document}
\draft



\title{ 
\vglue -0.5cm
 \hfill{\small IFT-P.075/2001} \\
\hfill{\small December 2001}\\
\hfill{\small hep-ph/0112248 }\\
\vglue 0.5cm
Lepton masses in a supersymmetric 3-3-1 model}

\author{\bf J. C. Montero\footnote{E-mail address: 
montero@ift.unesp.br}, 
V. Pleitez\footnote{E-mail address: vicente@ift.unesp.br} 
and M. C. Rodriguez\footnote{E-mail address: mcr@ift.unesp.br
}}
\address{
Instituto de F\'\i sica Te\'orica\\
Universidade Estadual Paulista\\
Rua Pamplona, 145\\ 
01405-900-- S\~ao Paulo, SP\\
Brazil}
\date{\today}

\maketitle
\begin{abstract}
We consider the mass generation for both charginos and neutralinos in a
3-3-1 supersymmetric model. We show that $R$-parity breaking interactions 
leave the electron and one of the neutrinos massless at the tree level. 
However the same interactions induce masses for these particles at the 1-loop 
level. Unlike the similar situation in the minimal supersymmetric standard model
the masses of the neutralinos are related to the masses of the charginos.

\end{abstract}

\pacs{
PACS number(s): 12.60.Jv, 14.60.-z, 14.60.Pq.}

\begin{multicols}{2}
\narrowtext

\section{Introduction}
\label{sec:intro}

The generation of neutrino masses is an important issue in any realistic 
extension of the standard model. In general, the values of these masses (of the 
order of, or less than, 1 eV) that are needed to explain all neutrino 
oscillation data~\cite{nusat,nussol,lsnd} are not enough to put strong 
constraints on model building. It means that several models can induce neutrino
masses and mixing compatible with experimental data. So, instead of 
proposing models built just to explain the neutrino properties, it is more
useful to consider what are the neutrino masses that are predicted in any
particular model which has motivation other than the explanation of neutrino
physics.   
For instance, the 3-3-1 model was proposed as a possible symmetry on the
lightest lepton sector $(\nu_e,e^-,e^+)_L$~\cite{331}. 
Once assumed that symmetry it has to be implemented in the rest of the leptons 
and also in the quark sector. Like in the standard model if we do not introduce
right-handed neutrinos and/or violation of the total lepton number neutrinos 
remain massless at any order in perturbation theory. In this vain it has
been done some effort to produce neutrino masses in the context of that 
3-3-1 model and some of its extensions~\cite{331nm}.

In this work we consider the generation of neutrino masses in a supersymmetric
3-3-1 model with broken $R$ parity. We show that, as an effect of the mixing 
among all leptons of the same charge, at the tree level only one 
charged lepton and one neutrino remain massless but they gain mass 
through radiative corrections.
In order to compare this model we do the same calculations in the context of the
minimal supersymmetric standard model with $R$ broken parity also.
In both cases we are not assuming that sneutrinos gain non-vanishing vacuum
expectation values (VEVs) i.e., the only non-zero VEVs are those of the scalars
of the non-supersymmetric models.

The outline of this work is as follows. In Sec.~\ref{sec:mssm} we review the
origin of the lepton masses in the minimal supersymmetric
standard model context under the same assumptions
that we will use in the case of the 3-3-1 supersymmetric model. 
In Sec.~\ref{sec:331s} we consider the supersymmetric version of a 3-3-1 model 
which has only three triplets of Higgs 
scalars. We explicitly show that leptons gain mass only as a consequence of 
their mixing with gauginos and higgsinos. 
Our conclusions are found in the last section.

\section{Neutrino masses in the MSSM}
\label{sec:mssm}

Let us consider in this section the lepton masses in the minimal supersymmetric
standard model (MSSM)~\cite{mssm}. In this model the interactions are written in
terms of the left-handed (right-handed) $\hat{L}\sim({\bf2},-1)$ 
($\hat{l}^c\sim({\bf1},2)$) leptons, left-handed (right-handed) quarks 
$\hat{Q}\sim({\bf2},1/3)$ $(\hat{u}^c\sim({\bf1},-4/3),
\hat{d}^c\sim({\bf1},2/3))$; and the Higgs doublets
$\hat{H}_{1}\sim({\bf2},-1),\hat{H}_{2}\sim({\bf2},1)$. With those multiplets  
the superpotential that conserves $R$-parity is given by
$W_{2RC}+W_{3RC}+\bar{W}_{2RC}+\bar{W}_{3RC}$, where 
\begin{eqnarray}
W_{2RC}&=&\mu\epsilon\hat{H}_1\hat{H}_2,\nonumber \\ 
W_{3RC}&=&\epsilon\hat{L}_af^l_{ab}\hat{H}_1\hat{l}^c_b+
\epsilon\hat{Q}_if^u_{ij}\hat{H}_2\hat{u}^c_j +
\epsilon\hat{Q}_if^d_{ij} \hat{H}_1\hat{d}^c_j,
\label{mssmrpc}
\end{eqnarray}
while the $R$-parity violating terms are given by $W_{2RV}+W_{3RV}+
\bar{W}_{2RV}+\bar{W}_{3RV}$, where
\begin{eqnarray}
W_{2RV}&=&\mu_{0a}\epsilon \hat{L}_a\hat{H}_2,\nonumber \\
W_{3RV}&=&\epsilon\hat{L}_a\lambda_{abc}\hat{L}_b\hat{l^c}_c+
\epsilon\hat{L}_a \lambda^\prime_{aij}\hat{Q}_i\hat{d}^c_j
+ \hat{u}^c_i \lambda^{\prime\prime}_{ijk}\hat{d}^c_j\hat{d}^c_k,
\label{mssmrpv}
\end{eqnarray}
and we have suppressed $SU(2)$ indices, $\epsilon$ is the
antisymmetric $SU(2)$ tensor. Above, and below in the following, the subindices
$a,b,c$ run over the lepton generations $e,\mu,\tau$ but a superscript $^c$
indicates charge conjugation; $i,j,k=1,2,3$ denote quark generations.
We have also to add the soft terms that break the supersymmetry:

\end{multicols}
\hspace{-0.5cm}
\rule{8.7cm}{0.1mm}\rule{0.1mm}{2mm}
\widetext

\begin{eqnarray}
{\cal L}_{\rm soft}&=&-\frac{1}{2}\left(\sum_{p=1}^{3}
m_{\lambda}\lambda^p_A \lambda^p_A+
m'\lambda_B\lambda_B+H.c.\right) 
- M^2_L \tilde{L}^{\dagger}\tilde{L}-
M^2_l \tilde{l^c}^{\dagger}\tilde{l^c}-
M^2_Q \tilde{Q}^{\dagger}\tilde{Q}-M^2_u \tilde{u^c}^{\dagger}\tilde{u^c}
\nonumber \\ &-&
M^2_d \tilde{d^c}^{\dagger}\tilde{d^c}-M^2_1 \tilde{H_1}^{\dagger}\tilde{H_1}-
M^2_2 \tilde{H_2}^{\dagger}\tilde{H_2} 
- \left[ A_L H_1 \tilde{L} \tilde{l^c}+A_U H_2 \tilde{Q} \tilde{u^c}+
A_D H_1 \tilde{Q} \tilde{d^c} \right.
\nonumber \\ &+& \left. M^2_{12}H_1H_2+BH_2L + C_1 \tilde{L}\tilde{L}\tilde{l^c}+
C_2 \tilde{L}\tilde{Q}\tilde{d^c}+
C_3 \tilde{u^c}\tilde{d^c}\tilde{d^c}+H.c. \right],
\label{softmssm}
\end{eqnarray}

\hspace{9.1cm}
\rule{-2mm}{0.1mm}\rule{8.7cm}{0.1mm}

\begin{multicols}{2}
\narrowtext

\noindent where $p$ is an $SU(2)$ index and $\lambda_A,\lambda_B$ are the
supersymmetric partners of the respective gauge vector bosons but we have
omitted generation indices and the gluino-mass terms.

With the interactions in
Eq.~(\ref{mssmrpc}) it is possible to give mass to all charged fermions in the
model (see below) but neutrinos remain massless.  
Hence, we must introduce $R$-parity violating term like those in
Eq.~(\ref{mssmrpv}). Some of the coupling constants in that expression should be
set to zero in order to avoid a too fast proton decay. Defining the basis
$\Psi^0_{MSSM}=(\nu_e,\nu_\mu,\nu_\tau,-i\lambda^3_A,-i\lambda_B,\tilde{H}^0_1,
\tilde{H}^0_2)^T$, the mass term is $-(1/2)[\Psi^{0T}_{MSSM}Y^{0}_{MSSM} 
\Psi^{0}+H.c.]$, where $Y^{0}_{MSSM}$ is the mass matrix 

\end{multicols}
\hspace{-0.5cm}
\rule{8.7cm}{0.1mm}\rule{0.1mm}{2mm}
\widetext
\begin{equation}
Y^0_{MSSM}= \left( \begin{array}{ccccccc}
0&  0& 0  &  0  & 0              & 0& -\mu_{0e}\\
0&  0& 0  &  0  & 0              & 0& -\mu_{0\mu} \\
0&  0& 0  &  0  & 0              & 0& -\mu_{0\tau} \\
0 & 0 & 0 &  m_{\lambda}  & 0        &  M_Z\,s_\beta c_W& -M_Z\,c_\beta c_W\\ 
0& 0& 0& 0 &m'                  & M_Z\,s_\beta s_W& -M_Z\,c_\beta s_W \\
0& 0& 0& M_Z\,s_\beta c_W&       M_Z\,s_\beta s_W& 0& \mu \\
-\mu_{0e} &-\mu_{0\mu}  &  -\mu_{0\tau}&  -M_Z\,c_\beta c_W& 
-M_Z\,c_\beta s_W& \mu& 0 
\end{array}
\right),
\label{mssm}
\end{equation}

\hspace{9.1cm}
\rule{-2mm}{0.1mm}\rule{8.7cm}{0.1mm}

\begin{multicols}{2}
\narrowtext
\noindent with $s_\beta=\sin\beta$, $s_W=\sin\theta_W$, etc are defined as
$\tan\beta=v_2/v_1$ and $\theta_W$ is the weak mixing angle. The matrix in 
Eq.~(\ref{mssm}) is generated only by the two usual vacuum expectation values 
of the two scalars and by the $R$-parity breaking terms $\mu_{0a}$. The mass 
matrix is similar to that in Ref.~\cite{hall,banks,rv1} but we have included 
the three neutrinos and we are neither assuming that sneutrinos gain nonzero 
vacuum expectation values nor have introduced sterile neutrinos like in 
Ref.~\cite{fb}. The mass matrix in Eq.~(\ref{mssm}) has two zero eigenvalues: 
it has determinant equal to zero and its secular equation which give the 
eigenvalues, $x$, has the form $x^2$ times a polynomial of five degree;  thus
there are two neutrinos $\nu_{1,2}$, which are massless at the tree level. Using
$\tan\beta=1$ and $M_Z=91.187$ GeV, $s^2_W=0.223$, $\mu_{0e}=\mu_{0\mu}=0$,
$\mu_{0\tau}=10^{-4}$ GeV (this value is consistent with that of 
Ref.~\cite{banks}), $\mu=100$ GeV, $m=250$ GeV, $m'=-200$ GeV, we obtain
besides the two massless neutrinos a massive one with
$m_{\nu_3}=-3\times10^{-3}$ eV, and four heavy neutralinos with masses $267.40,
-199.99, -117.40, 100.0$ GeV. These zero eigenvalues are a product of the 
matrix structure in Eq.~(\ref{mssm}) and there is not a symmetry to protect the 
neutrinos to gain mass by radiative corrections. On the other hand, if 
$\mu_{0a}=0$, $a=e,\mu,\tau$, all neutrinos remain massless at the tree level. 
In this case it is the $R$-parity and total lepton number conservation 
that protect neutrinos of gain masses. 
The neutralino masses above are consistent with those of Ref.~\cite{banks}: two states are massless
and the other ones have masses of the order $O(M_Z)$. More realistic neutrino
masses require radiative corrections~\cite{rv1,rnm,rv2,marta}.
Here we will only consider the neutrino masses generated by radiative
corrections arisen from the interactions given in Eqs.~(\ref{mssmrpc}) and
(\ref{mssmrpv}) and only two VEVs. We have in this case the interactions
\begin{eqnarray}
&&-\frac{\lambda_{abc}}{3}\left( \bar{\nu}_{aL}l_{bR}\tilde{l}_c+
\bar{\nu}^c_{aR}l^c_{bL}\tilde{l}^*_c\right)
\nonumber \\ &-&\frac{\lambda^\prime_{aij}}{3}\left( \bar{\nu}_{aL}d_{iR}\tilde{d}_j+
\bar{\nu}^c_{aR}d^c_{iL}\tilde{d}^*_j\right)+H.c.,
\label{nusint1}
\end{eqnarray}
and the 1-loop diagrams like those in Ref.~\cite{hall} arise. Notice however
that if we introduce a discrete symmetry (called $Z^\prime_3$ later on), 
$\hat{L}_{e,\mu}\to-\hat{L}_{e,\mu}$, and all other fields are even under
this transformation, we have that 
\begin{eqnarray}
\mu_{0e}=\mu_{0\mu}=0;\; \nonumber \\
\lambda_{ebc}=0,\;(b,c=\mu,\tau);\;
\lambda_{\mu bc}=0,\;\, (b,c=e,\tau); \nonumber \\
\lambda^\prime_{eij}=\lambda^\prime_{\mu ij}=0,\;\,(i,j=1,2,3);
\label{yuju}
\end{eqnarray}
and the $\nu_e,\nu_\mu$ neutrinos will remain massless at all order
in perturbation theory. It is also possible to choose the symmetry such as
$L_{e,\tau} \to- L_{e,\tau}$ while all other fields remain invariant. 
In this case we have that $\nu_e$ and $\nu_\tau$ remain massless. However, if no
discrete symmetry is imposed neutrinos gain mass through 1-loop effect like in
Ref.~\cite{hall}.

Next, let us consider the charged sector. 
With the interactions in Eq.~(\ref{mssmrpc}) it is possible to give mass to all
charged fermions. Denoting
\begin{equation}
\begin{array}{c}
\phi^+_{MSSM}=(e^c,\mu^c,\tau^c,-i\lambda^+_W,  
\tilde{H}^+_2)^T,\\
\phi^-_{MSSM}=(e,\mu,\tau,-i\lambda^-_W,
\tilde{H}^{-}_1)^T,
\end{array}
\label{cbasismssm}
\end{equation}
where all the fermionic fields are still Weyl spinors, we can define
$\Psi^{\pm}_{MSSM}=(\phi^+_{MSSM},\, \phi^-_{MSSM})^T$, and the mass term   
$-(1/2)[\Psi^{\pm T}_{MSSM}Y^\pm_{MSSM}\Psi^{\pm}_{MSSM}+H.c.]$ where $Y^\pm$ is 
the mass matrix given by: 
\begin{equation}
Y^{\pm}_{MSSM}= \left( \begin{array}{cc}
0  & X^T_{MSSM}\\
X_{MSSM}  & 0
\end{array}
\right),
\label{ypmmssm}
\end{equation}
with
\end{multicols}
\hspace{-0.5cm}
\rule{8.7cm}{0.1mm}\rule{0.1mm}{2mm}
\widetext
\begin{equation}
X_{MSSM}= \left( \begin{array}{ccccc}
-f^l_{ee}v_1    & -f^l_{e\mu}v_1     & -f^l_{e\tau}v_1      &0 & 0  \\
-f^l_{e\mu}v_1  & -f^l_{\mu\mu}v_1   &  -f^l_{\mu\tau}v_1   & 0& 0 \\ 
-f^l_{e\tau}v_1 & -f^l_{\mu\tau}v_1  & -f^l_{\tau\tau}v_1   & 0& 0\\ 
0            & 0               & 0      & m_{\lambda} & \sqrt{2}M_Wc_\beta \\
\mu_{0e} & \mu_{0\mu} & \mu_{0\tau}  & \sqrt{2}M_Ws_\beta  & \mu 
\end{array}
\right).
\label{clmmmssm}
\end{equation}

\hspace{9.1cm}
\rule{-2mm}{0.1mm}\rule{8.7cm}{0.1mm}

\begin{multicols}{2}
\narrowtext

With $f^l_{ee}=2.7\cdot10^{-4}$, $f^l_{\mu\mu}=3.9\cdot10^{-3}$,
$f^l_{\tau\tau}=1.6\cdot10^{-2}$, $f^l_{e\mu}=f^l_{e\tau}=f^l_{\mu\tau}=10^{-7}$
we obtain from Eq.~(\ref{clmmmssm}) the masses $0.0005,0.105,1.777$ (in GeV) for
the usual leptons, and 4.3 and 81 TeV for the charginos. 
We see by comparing Eq.~(\ref{mssm}) with Eq.~(\ref{clmmmssm}) that there is 
no relation between the charged lepton masses and the neutralino
masses. Notice also that all charged leptons gain masses at the tree level. 
We will not consider this model (or some of its extensions) further since it 
has been well studied in literature~\cite{hall,banks,rv1,rnm,rv2,marta}. 

\section{A supersymmetric 3-3-1 model}
\label{sec:331s}

In the nonsupersymmetric 3-3-1 model~\cite{331} the fermionic representation
content is as follows: left-handed leptons $L=(\nu_a,l_a,l^c_a)_L
\sim({\bf1},{\bf3},0)$, $a=e,\mu,\tau$; left-handed quarks 
$Q_{1L}=(u_1,d_1,J)\sim({\bf3},{\bf3},2/3)$,
$Q_{\alpha L}=(d_\alpha,u_\alpha,j_\beta) 
\sim({\bf3},{\bf3}^*,-1/3)$, $\alpha=2,3$,
$\beta=1,2$; and in the right-handed components we have 
$u^c_i,d^c_i,\,i=1,2,3$, that transform as in the SM, and the exotic quarks 
$J^c\sim({\bf3}^*,{\bf1},-5/3),j_\beta\sim({\bf3}^*,{\bf1},4/3)$.
The minimal scalar
representation content is formed by three scalar triplets: 
$\eta\sim({\bf1},{\bf3},0)=(\eta^0,\eta^-_1,\eta^+_2)^T$;
$\rho\sim({\bf1},{\bf3},+1)=(\rho^+, \rho^0, 
\rho^{++})^T$ and $\chi\sim
({\bf1},{\bf3},-1)=(\chi^-,\chi^{--},\chi^0)^T$, and one scalar sextet
$S\sim({\bf1},{\bf6},0)$.  
We can avoid the introduction of the sextet by adding a charged lepton
transforming as a singlet~\cite{dma,lepmass1}. Notwithstanding, here we will
omit both the sextet and the exotic lepton. A see-saw-type mechanism will be
implemented by the mixing with supersymmetric partners, higgsinos or
gauginos. The complete
set of fields in the 3-3-1 supersymmetric model has been given in 
Refs.~\cite{331s,mcr}. We will denote, like in the previous 
section, the respective superfields as $\hat{L}$ and so on.
We recall that in the nonsupersymmetric 3-3-1 model with only the three
triplets the charged lepton masses are not yet the physical ones: $0,m,-m$. 

We will show how, in the present model supersymmetry and the
$R$-violating interactions give the correct masses to 
$e,\mu$ and $\tau$, even without a sextet or  
the charged lepton singlet. We have the 
higgsinos $\tilde\eta,\tilde\rho,\tilde\chi$ and their respective primed 
fields which have the same charge assignment of the triplets 
$\eta,\rho$ and $\chi$, for details see Ref.~\cite{331s}.

Due to the fact that in the supersymmetric model we have the gauginos and
higgsinos, (for details on the lagrangian of the model see \cite{331s}), when 
the $R$-parity is broken we have in analogy with the MSSM, but with important
differences, a mixture between the usual leptons and the gauginos and higgsinos. 

One part of the superpotential is given by $W_2+\bar{W}_2$ where
\begin{eqnarray}
W_2=\mu_{0a}\hat{L}_a\hat{\eta}^\prime+
\mu_\eta\hat{\eta}\hat{\eta}^\prime+
\mu_\rho\hat{\rho}\hat{\rho}^\prime+\mu_\chi\hat{\chi}\hat{\chi}^\prime,
\label{e2}
\end{eqnarray}
$a=e,\mu,\tau$; and $W_3+\bar{W}_3$ where
\begin{eqnarray}
W_3&=&\lambda_{1abc}\epsilon\hat{L}_a\hat{L}_b\hat{L}_c+
\lambda_{2ab}\epsilon\hat{L}_a\hat{L}_b\hat{\eta}+
\lambda_{3a}\hat{L}_a\hat{\rho}\hat{\chi}\nonumber \\ &+&
f_1\epsilon\hat{\eta}\hat{\rho}\hat{\chi} +
f^\prime_1\epsilon\hat{\eta}^\prime\hat{\rho}^\prime\hat{\chi}^\prime
+\lambda^\prime_{\alpha ai} \hat{Q}_{\alpha} \hat{L}_{a} 
\hat{d^c}_{i}\nonumber \\ &+&
\lambda^{\prime\prime}_{ijk}\hat{u^c_i}\hat{d^c_j}\hat{d^c_k} +
\lambda^{\prime\prime\prime}_{ij\beta}\hat{u^c_i}\hat{u^c_j}\hat{j^c_\beta}
\nonumber \\ &+&
\lambda^{\prime\prime\prime\prime}_{i\beta}\hat{d^c_i}\hat{J^c}\hat{j^c_\beta}+
\kappa_{1i} \hat{Q}_{1} \hat{\eta}^{\prime} \hat{u}^{c}_{i} \nonumber \\ &+&
\kappa_{2i} \hat{Q}_{1} \hat{\rho}^{\prime} \hat{d}^{c}_{i}+
\kappa_{3} \hat{Q}_{1} \hat{\chi}^{\prime} \hat{J}^{c}+
\kappa_{4\alpha i} \hat{Q}_{\alpha} \hat{\eta} \hat{d}^{c}_{i} \nonumber \\
&+&
\kappa_{5\alpha i} \hat{Q}_{\alpha} \hat{\rho} \hat{u}^{c}_{i}+
\kappa_{6\alpha \beta} \hat{Q}_{\alpha} \hat{\chi} \hat{j}^{c}_{\beta},
\label{e22}
\end{eqnarray}
with $\epsilon$ the completely antisymmetric tensor of $SU(3)$ but we have
omitted the respective indices; the generation indices are as follows:
$a,b,c=e,\mu,\tau$ and $i,j,k=1,2,3$. 

The gaugino masses come from 
the soft-terms shown in the Appendix~\ref{sec:a1}, Eq.~(\ref{ess}). 
The $\mu_0$, $\lambda_1,\lambda_3,\lambda^\prime,
\lambda^{\prime\prime},\lambda^{\prime\prime\prime}$ and
$\lambda^{\prime\prime\prime\prime}$ terms break the $R$-parity defined in this
model as 
$R=(-1)^{3{\cal F}+2{\rm S}}$ where ${\cal F}={\rm B+L}$, B(L) is the 
baryon (total lepton) number; S is  the spin. The $\lambda_2$ term of the 
superpotential $W_3$ implies interactions like (see Eq.~(\ref{nusint}) below)
$\bar\nu_{aL}\tilde{\eta}^-_{2R}\tilde{l}^*_b- 
\bar{\nu}_{aL}\tilde{\eta}^+_{1R}\tilde{l}_b$ and we have also the interactions 
\begin{equation}
{\cal L}_\eta=\int d^4\,\theta \hat{\bar{\eta}}e^{2g\hat{V}}\hat{\eta},
\label{e5}
\end{equation}
where $\hat{V}$ is the superfield related to the $V^{a}$ gauge boson of
$SU(3)_L$. This interaction mixes higgsinos with gauginos as showed 
in Ref.~\cite{mcr}. 

The parameters $\mu_\eta$ and $\mu_\rho$ are the equivalent of the $\mu$
parameter in the MSSM~\cite{mssm}. The terms proportional to 
$\lambda_2$ and $\mu_\chi$ have no equivalent in the MSSM. The $\lambda^\prime$
and $\lambda^{\prime\prime}$ coupling constants are constrained by the proton
decay such that~\cite{pal}
\begin{equation}
\lambda^{\prime\prime}_{11j}\lambda^\prime_{a2j}<10^{-24},
\label{lcon}
\end{equation}
assuming the superpartner masses in the range of 1 TeV.     

\subsection{Charged lepton masses}
\label{subsec:clm}

In this model there are interactions like
\begin{eqnarray}
&&-\frac{\lambda_{3a}}{3}\left[ \omega
(l_a\tilde{\rho}^++\bar{l}_a\overline{\tilde{\rho}^{+}}) +
u(l^c_a\tilde{\chi}^-+\bar{l^c}_a\overline{\tilde{\chi}^{-}})
\right] \nonumber \\
&-&\frac{1}{2}\,\mu_{0a}[l\tilde{\eta}_1^{\prime+}+
\bar{l}\overline{\tilde{\eta}_1^{\prime+}}+
l^c\tilde{\eta}_2^{\prime-} +
\bar{l^c}\overline{\tilde{\eta}_2^{\prime-}}],
\label{newint}
\end{eqnarray}
which imply a general mixture in both neutral and charged sectors.
Let us first considered the charged lepton masses.
Denoting
\begin{equation}
\begin{array}{c}
\phi^+=(e^c,\mu^c,\tau^c,-i\lambda^+_W,-i\lambda^+_V, 
\tilde{\eta}^{\prime+}_1,\tilde{\eta}^+_2,
\tilde{\rho}^+,\tilde{\chi}^{\prime+})^T,\\
\phi^-=(e,\mu,\tau,-i\lambda^-_W,-i\lambda^-_V, 
\tilde{\eta}^{-}_1,\tilde{\eta}^{\prime -}_2,
\tilde{\rho}^{\prime -},\tilde{\chi}^{-})^T,
\end{array}
\label{cbasis}
\end{equation}
where all the fermionic fields are still Weyl spinors, we can also, as before, 
define $\Psi^{\pm}=(\phi^+ \phi^-)^T$, and the mass term 
$-(1/2)[  
\Psi^{\pm T}Y^\pm\Psi^{\pm}+H.c.]$
where $Y^\pm$ is given by: 
\begin{equation}
Y^{\pm}= \left( \begin{array}{cc}
0  & X^T \\
X  & 0
\end{array}
\right),
\label{ypm}
\end{equation}
with
\end{multicols}
\hspace{-0.5cm}
\rule{8.7cm}{0.1mm}\rule{0.1mm}{2mm}
\widetext
\begin{equation}
X= \left( \begin{array}{ccccccccc}
0&- \frac{\lambda_{2e \mu}}{3}v&- \frac{ \lambda_{2e \tau}}{3}v& 0& 0&- 
\frac{\mu_{0e}}{2}& 0&- \frac{\lambda_{3e}}{3}w& 0\\
\frac{ \lambda_{2e \mu}}{3}v& 0&- \frac{ \lambda_{2\mu \tau}}{3}v& 0& 0&- 
\frac{\mu_{0\mu}}{2}& 0&- \frac{\lambda_{3 \mu}}{3}w& 0\\
\frac{ \lambda_{2e \tau}}{3}v& \frac{ \lambda_{2\mu \tau}}{3}v& 0& 0& 0&- 
\frac{\mu_{0\tau}}{2}& 0&- \frac{\lambda_{3 \tau}}{3}w& 0\\
0& 0& 0& m_{\lambda}& 0&- gv^{\prime}& 0& gu& 0\\
0& 0& 0& 0& m_{\lambda}& 0& gv& 0&- gw^{\prime}\\
0& 0& 0& gv& 0&- \frac{\mu_{\eta}}{2}& 0& \frac{f_{1}w}{3}& 0\\
- \frac{\mu_{0e}}{2}&- \frac{\mu_{0\mu}}{2}&- \frac{\mu_{0\tau}}{2}& 0&- 
gv^{\prime}& 0&- \frac{\mu_{\eta}}{2}& 0&- \frac{f^{\prime}_{1}u^{\prime}}{3}\\
0& 0& 0&- gu^{\prime}& 0& \frac{f^{\prime}_{1}w^{\prime}}{3}& 0&- 
\frac{\mu_{\rho}}{2}& 0\\
- \frac{\lambda_{3e}}{3}u&- \frac{\lambda_{3 \mu}}{3}u&- 
\frac{\lambda_{3 \tau}}{3}u& 0& gw& 0&- \frac{f_{1}u}{3}& 0&- 
\frac{\mu_{\chi}}{2}
\end{array}
\right),
\label{clmm}
\end{equation}

\hspace{9.1cm}
\rule{-2mm}{0.1mm}\rule{8.7cm}{0.1mm}

\begin{multicols}{2}
\narrowtext
where we have defined 

\begin{eqnarray}
v=\frac{v_\eta}{\sqrt2},\;u=\frac{v_\rho}{\sqrt2},\;w=\frac{v_\chi}{\sqrt2},
\nonumber \\
v^\prime=\frac{v^\prime_\eta}{\sqrt2},\; u^\prime=\frac{v^\prime_\rho}{\sqrt2},
\;w^\prime=\frac{v^\prime_\chi}{\sqrt2}.
\label{vevs}
\end{eqnarray}

The chargino mass matrix $Y^\pm$ is diagonalized using two unitary matrices, $D$
and $E$, defined by
\begin{eqnarray}
\tilde{ \chi}^{+}_{i}=D_{ij} \Psi^{+}_{j}, \,\
\tilde{ \chi}^{-}_{i}=E_{ij} \Psi^{-}_{j}, \,\ i,j=1, \cdots , 9,
\label{2sc} 
\end{eqnarray}
($D$ and $E$ sometimes are denoted, in non-supersymmetric theories, by $U^l_R$
and $U^l_L$, respectively). Then we can write the diagonal mass matrix as 
\begin{equation}
M_{SCM}=E^{*}XD^{-1}.
\label{m1}
\end{equation}
To determine $E$ and $D$, we note that
\begin{equation}
M^{2}_{SCM}=DX^T \cdot XD^{-1}=E^{*}X \cdot X^T(E^{*})^{-1},
\label{m2}
\end{equation}
and define the following Dirac spinors:
\begin{eqnarray}
\Psi(\tilde{ \chi}^{+}_{i})= \left( \begin{array}{cc}
             \tilde{ \chi}^{+}_{i} &
	     \bar{ \tilde{\chi}}^{-}_{i} 
\end{array} \right)^T, \,\
\Psi^{c}(\tilde{ \chi}^{-}_{i})= \left( \begin{array}{cc}
             \tilde{ \chi}^{-}_{i} &
	     \bar{ \tilde{\chi}}^{+}_{i} 
\end{array} \right)^T,
\label{emasssim}
\end{eqnarray}
where $\tilde{ \chi}^{+}_{i}$ is the particle and $\tilde{ \chi}^{-}_{i}$ 
is the anti-particle~\cite{mssm,mcr}.

We have obtained the following masses (in GeV) for the charged sector:
\begin{equation}
3186.05, 3001.12, 584.85, 282.30,
 204.55, 149.41, 
\label{cm}
\end{equation}
and the masses for the usual leptons (in GeV) $m_e=0$, $m_\mu=0.1052$ and 
$m_\tau=1.777$. These values have been obtained by
using  the following values for the dimensionless parameters
\begin{mathletters}
\label{valpara}
\begin{eqnarray}
\lambda_{2e\mu}=0.001,\;\lambda_{2e\tau}=0.001,\;\lambda_{2\mu\tau}=0.393,
\nonumber \\
\lambda_{3e}=0.0001,\; \lambda_{3\mu}=1.0,\;\lambda_{3\tau}=1.0,
\label{lambdas}
\end{eqnarray}
\begin{equation}
f_{1}=0.254,\;f^{\prime}_{1}=1.0,
\label{efes}
\end{equation}
and for the mass dimension parameters (in GeV) we have used:
\begin{equation}
\mu_{0e}=\mu_{0\mu}=0.0,\quad \mu_{0\tau}=10^{-6},
\label{mus0}
\end{equation}
\begin{equation}
\mu_{ \eta}=300,\; \mu_{ \rho}=500, \;\mu_{ \chi}=700,\;  m_{\lambda}=3000. 
\label{para}
\end{equation}
\end{mathletters}
We also use the constraint $V^{2}_{ \eta}+V^{2}_{ \rho}=(246\;{\rm GeV})^2$ 
coming from $M_{W}$, where, we have defined 
$V^{2}_{ \eta}=v^{2}_{ \eta}+v^{ \prime 2}_{ \eta}$ and 
$V^{2}_{ \rho}=v^{2}_{ \rho}+v^{ \prime 2}_{ \rho}$.
Assuming that $v_{ \eta}=20$ GeV, $v^{ \prime}_{ \eta}=
v^{ \prime}_{ \rho}=1$ GeV, and 
$2v_{ \chi}=v^{ \prime}_{ \chi}=2$ TeV, the
value of $v_\rho$ is fixed by the constraint above. 

\begin{figure}
\begin{center}
\includegraphics[width=8cm]{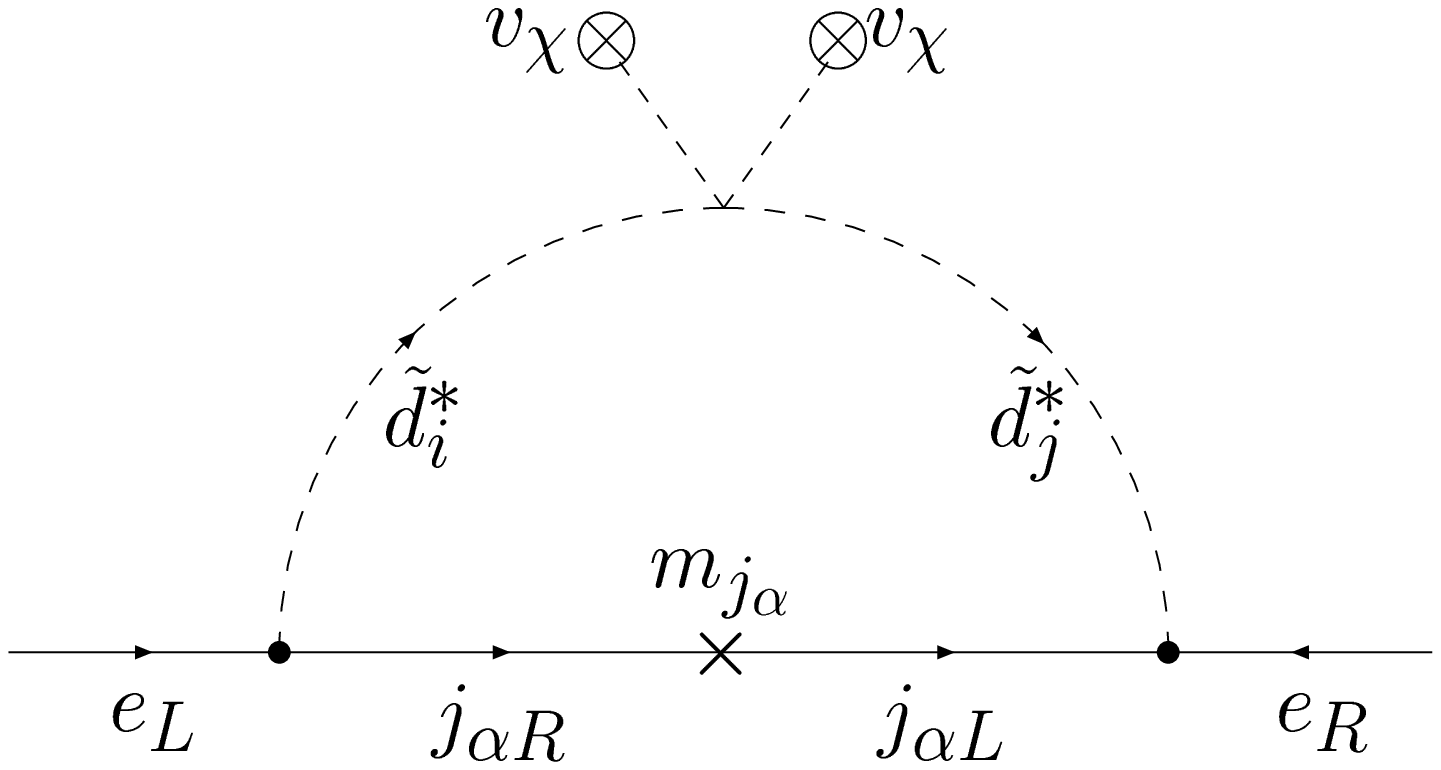}
\end{center}
\caption{Diagram generating the electron mass. There are also a contribution
with $v_\chi\to v_{\chi'}$. The left- and right- side vertices are proportional
to $\lambda^\prime_{\alpha e i}/3$ and $\lambda^\prime_{\alpha' e j}/3$,
respectively. 
}
\label{fig1}
\end{figure}

Notice, from Eq.~(\ref{cm}), that the electron is massless at the tree level. 
This is again a result of the structure of the mass matrix in 
Eq.~(\ref{clmmmssm}) and there is not a symmetry that protects the electron 
to get a mass by loop corrections. Hence,
it can gain mass trough radiative corrections like that shown in 
Fig.~\ref{fig1}. The interactions of the leptons with the sleptons written in 
term of Dirac fermions (although we are using the same notation) are given by 
(and the respective hermitian conjugate)

\end{multicols}
\hspace{-0.5cm}
\rule{8.7cm}{0.1mm}\rule{0.1mm}{2mm}
\widetext
\begin{eqnarray}
{\cal L}^l&=&
\frac{\lambda_{2ab}}{3}\left[\,\overline{\tilde{\eta}^-_{2R}}\,
(l_{bL}\tilde{\nu}_{la}-l_{aL}\tilde{\nu}_{lb})+
\overline{\tilde{\eta}^0_R}[(l_{aL}\tilde{l}^*_b-l_{bL}\tilde{l}^*_a)
+(l^c_{bL}\tilde{l}_a- l^c_{aL}\tilde{l}_b)]+
\overline{\tilde{\eta}^+_{1R}}\,
(l^c_{aL}\tilde{\nu}_{bL}-l^c_{bL}\tilde{\nu}_{aL})\right] \nonumber \\ &+&
\frac{\lambda_{3a}}{3}\left[
-\overline{\tilde{\rho}^{--}_R}l_{aL}\chi^-+
\bar{\tilde\rho}^-_R\,( l_{aL}\chi^0- l^c_{aL}\chi^{--})-
\overline{\tilde{\chi}^+_R}\,(l_{aL}\rho^{++}-l^c_{al}\rho^0)-
\overline{\tilde{\chi}^{++}_R}l^c_{aL}\rho^++
\overline{\tilde{\rho}^0_R}l^c_{aL}\chi^{-}
+\overline{\tilde{\chi}^0_R}l_{aL}\rho^+\right]\nonumber \\ &+&
\frac{\lambda^\prime_{\alpha ai}}{2}
\left[(\bar{u}_{\alpha R}l_{aL}+ \bar{j}_{\alpha R}l^c_{aL} )\tilde{d}^*_i
+\bar{d}^c_{iR}\,( l_{aL}\tilde{u}_\alpha + l^c_{aL}\tilde{j}_\alpha)\right].
\label{clint}
\end{eqnarray}

\hspace{9.1cm}
\rule{-2mm}{0.1mm}\rule{8.7cm}{0.1mm}

\begin{multicols}{2}
\narrowtext

The $\lambda^\prime$ interactions generate the low vertices in Fig.~\ref{fig1}. 
On the other hand, the interactions between the squarks, sleptons and scalars, 
see the Appendix~\ref{sec:a1}, are given by the scalar potential. The soft part contributes
only through the trilinear interactions
\begin{equation}
V_{soft}=
\epsilon_{1ab}(\tilde{l}_a\tilde{l}^*_b- \tilde{l}^*_a\tilde{l}_b)\eta^0,
\label{softnew}
\end{equation}
while the $D$-terms have only quartic interactions
\begin{equation}
V_D=\frac{g^2}{4}\sum_i(X^0_iX^0_i+X^{\prime 0}_aX^{\prime
0}_i)\sum_a\left(\tilde{l}_a\,\tilde{l}^*_a + \frac{1}{2}
\tilde{\nu}_a\,\tilde{\nu}^*_a\right),
\label{dt}
\end{equation}
where $X^0_i=\chi^0,\eta^0;X^{\prime 0}_i=\eta^{\prime 0},\chi^{\prime 0}$; and
$a=e,\mu,\tau$. From the $F$-terms we have contributions to both, the 
trilinear interactions
\begin{equation}
V_{3F}=\left( \frac{\mu_{0c}\lambda_{1abc}}{2}+\frac{\mu_\chi\lambda_{2ab}}{6}
\right)\left(\tilde{l_a}\tilde{l}^*_b -\tilde{l^*_a}
\tilde{l}_b\right)\eta^{\prime0}+H.c., 
\label{ft3}
\end{equation}
and the quartic ones
\end{multicols}
\hspace{-0.5cm}
\rule{8.7cm}{0.1mm}\rule{0.1mm}{2mm}
\widetext
\begin{eqnarray}
V_{4F}&=&\left(\frac{\lambda_{3d} \lambda_{1dab}}{3}+
\frac{f_1\lambda_{2ab}}{9}\right)
(\tilde{l}_a\tilde{l^*}_b-\tilde{l^*}_a\tilde{l}_b)\rho^0\chi^0+
\frac{4\lambda_{2ad}\lambda_{2db}}{9}\left( \tilde{l}^*_a
\tilde{l}_b+\tilde{l}_a\tilde{l^*}_b\right)\eta^0\eta^0\nonumber \\ &+&
\frac{\lambda_{3a}\lambda_{3b}}{9}\left(\tilde{l}_a\tilde{l}_b^*
+\tilde{\nu}_a\tilde{\nu}_b^*\right)\,
\sum_iX^0_iX^0_i,
\label{ft4}
\end{eqnarray}
\hspace{9.1cm}
\rule{-2mm}{0.1mm}\rule{8.7cm}{0.1mm}

\begin{multicols}{2}
\narrowtext
\noindent where  $X^0_i=\chi^0,\rho^0$ which
will contribute to the upper quartic vertex in Fig.~\ref{fig1}. Due to the 
interactions given in
Eq.~(\ref{clint})--(\ref{ft4}), we can generate
the appropriate mass to the electron. The dominant  
contributions, assuming the mass hierarchy $m_{\rm fermion}\ll m_{\rm scalar}$
where fermion means a fermion different from $j_{1,2}$ and scalar means
$\tilde{\nu},\tilde{l},H$ ($H$ denote the heaviest Higgs scalar) and using the
values of the masses and the parameters given in Eqs.~(\ref{cm}),
(\ref{lambdas}) and (\ref{para}) we obtain that the dominant 
contribution to the electron mass is, up to logarithmic corrections, 
\begin{equation}
m_e\propto \lambda^\prime_{\alpha ei}\lambda^\prime_{\alpha' ej} 
V^2_j V^2_b
(v^2_\chi+v^2_{\chi'})\,\frac{m_{j_\alpha}}{9m^2_{\tilde{b}}},
\label{emass}
\end{equation}
and with all the indices fixed, $V_j$ denotes mixing matrix elements in the two
dimension $j_{1,2}$ space, $V_b$ means the same but in the d-like squark sector. 
We obtain $m_e=0.0005$ GeV if $v_\chi$ and $v^\prime_\chi$ have the values
already giving above, and with $\lambda^\prime_{\alpha
ei}\lambda^\prime_{\alpha'ej}\approx 10^{-6}$, which imposes 
\begin{equation}
\frac{m_{j_\beta}}{m^2_{\tilde{b}}}\approx \frac{9\times 10^{-4}}{V^2_jV^2_b}
\,\,{\rm GeV}^{-1},
\label{r1}
\end{equation}
or $ m_{\tilde{b}}= 33.33\sqrt{m_{j_2}}V_jV_b$ GeV. Using $m_j\sim250(320)$
GeV~\cite{das}, we have $m_{\tilde{b}}\sim 526(596) V_jV_b$ GeV. With 
$V_jV_b\sim0.14(0.12)$ we obtain squarks masses of the order of
$m_{\tilde{b}}\sim75$ GeV~\cite{pdg}.  

\subsection{Neutral lepton masses}
\label{subsec:nlm}

Like in the case of the charged sector, the neutral lepton masses are given 
by the mixing among neutrinos, induced by the $\mu_{0a}$ term in Eq.~(\ref{e2}), 
and the neutral higgsinos and gauginos~\cite{mcr}, 
and also by $\lambda_2$ and $\lambda_3$ in Eq.~(\ref{e22}).
The first two terms in Eq.~(\ref{e2}) give the interactions between neutrinos 
and higgsinos:
\begin{equation}
\frac{1}{2}\,\mu_{0a}[\nu_a\tilde{\eta}^{\prime0}+
\bar{\nu}_a\overline{\tilde{\eta}^{\prime0}}]-
\frac{\mu_\eta}{2}\,[\tilde{\eta}^{\prime0}\tilde{\eta}^{0}+
\overline{\tilde{\eta}^{\prime0}} \,\ \overline{\tilde{\eta}^{0}}],
\label{e3}
\end{equation}
and from Eq.~(\ref{e22}) we have also the interactions
\begin{equation}
\frac{\lambda_{3a}}{3}
[w(\nu_a\tilde{\rho}^0+\bar{\nu}_a\overline{\tilde\rho^0})
+u(\nu_a\tilde{\chi}^0+\bar{\nu}_a\overline{\tilde{\chi}^0})].
\end{equation}

These interactions imply a mass term for the neutrinos and neutralinos. 
The mass term in the basis
\begin{equation}
\Psi^{0}= \left( 
\nu_{e} \nu_{\mu} \nu_{\tau}
-i \lambda^{3}_{A}
-i \lambda^{8}_{A}
-i \lambda_{B}
\tilde{\eta}^{0}
\tilde{\eta}^{ \prime 0}
\tilde{\rho}^{0}
\tilde{\rho}^{ \prime 0}
\tilde{\chi}^{0}
\tilde{\chi}^{ \prime 0}
\right)^T, 
\end{equation}
is given by $-(1/2)[ \left( \Psi^{0} 
\right)^T Y^{0} \Psi^{0}+H.c.] $ where
\end{multicols}
\hspace{-0.5cm}
\rule{8.7cm}{0.1mm}\rule{0.1mm}{2mm}
\widetext

\begin{equation}
Y^{0}= \left( \begin{array}{cccccccccccc}
0& 0& 0& 0& 0& 0& 0&- \frac{\mu_{0e}}{2}& \frac{ \lambda_{3 e}}{3}w& 0& 
\frac{ \lambda_{3e}}{3}u& 0\\
0& 0& 0& 0& 0& 0& 0&- \frac{\mu_{0\mu}}{2}& \frac{ \lambda_{3 \mu}}{3}w& 0& 
\frac{ \lambda_{3\mu}}{3}u& 0\\
0& 0& 0& 0& 0& 0& 0&- \frac{\mu_{0\tau}}{2}& \frac{ \lambda_{3 \tau}}{3}w& 0& 
\frac{ \lambda_{3 \tau}}{3}u& 0\\
0& 0& 0& m_{\lambda}& 0& 0& \frac{gv}{\sqrt{2}}&- 
\frac{gv^{\prime}}{\sqrt{2}}&- 
\frac{gu}{\sqrt{2}}& \frac{gu^{\prime}}{\sqrt{2}}& 0& 0\\
0& 0& 0& 0& m_{\lambda}& 0& \frac{gv}{\sqrt{6}}&- \frac{gv^{\prime}}{\sqrt{6}}& 
\frac{gu}{\sqrt{6}}&- \frac{gu^{\prime}}{\sqrt{6}}&- \frac{2}{\sqrt{6}} gw& 
\frac{2}{\sqrt{6}}gw^{\prime}\\
0& 0& 0& 0& 0& m^{\prime}& 0& 0& \frac{g^{\prime}u}{\sqrt{2}}&- 
\frac{g^{\prime}u^{\prime}}{\sqrt{2}}&-  \frac{g^{\prime}w}{\sqrt{2}}& 
\frac{g^{\prime}w^{\prime}}{\sqrt{2}}\\
0& 0& 0&  \frac{gv}{\sqrt{2}}& \frac{gv}{\sqrt{6}}& 0& 0&- \frac{\mu_{\eta}}{2}&- 
\frac{f_{1}w}{3}& 0& \frac{f_{1}u}{3}& 0\\
- \frac{\mu_{0e}}{2}&- \frac{\mu_{0\mu}}{2}&- \frac{\mu_{0\tau}}{2}&-
\frac{gv^{\prime}}{\sqrt{2}}&-  
\frac{gv^{\prime}}{\sqrt{6}}& 0&- \frac{\mu_{\eta}}{2}& 0& 0&- 
\frac{f^{\prime}_{1}w^{\prime}}{3}& 0& \frac{f^{\prime}_{1}u^{\prime}}{3}\\
\frac{ \lambda_{3e}}{3}w& \frac{ \lambda_{3 \mu}}{3}w& 
\frac{ \lambda_{3 \tau}}{3}w&- 
\frac{gu}{\sqrt{2}}& \frac{gu}{\sqrt{6}}& \frac{g^{\prime}u}{\sqrt{2}}&- 
\frac{f_{1}w}{3}& 0& 0&- \frac{ \mu_{\rho}}{2}&- \frac{f_{1}v}{3}& 0\\
0& 0& 0&  \frac{gu^{\prime}}{\sqrt{2}}&- \frac{gu^{\prime}}{\sqrt{6}}&- 
\frac{g^{\prime}u^{\prime}}{\sqrt{2}}& 0&- \frac{f^{\prime}_{1}w^{\prime}}{3}&- 
\frac{ \mu_{\rho}}{2}& 0& 0&- \frac{f^{\prime}_{1}v^{\prime}}{3}\\
\frac{ \lambda_{3e}}{3}u& \frac{ \lambda_{3\mu}}{3}u& 
\frac{ \lambda_{3 \tau}}{3}u& 0&
- \frac{2}{\sqrt{6}} gw&- \frac{g^{\prime}w}{\sqrt{2}}& \frac{f_{1}u}{3}& 0&- 
\frac{f_{1}v}{3}& 0& 0&- \frac{ \mu_{\chi}}{2}\\
0& 0& 0& 0& \frac{2}{\sqrt{6}} gw^{\prime}& \frac{g^{\prime}w^{\prime}}{\sqrt{2}}& 0& 
\frac{f^{\prime}_{1}u^{\prime}}{3}& 0&- \frac{f^{\prime}_{1}v^{\prime}}{3}&- 
\frac{ \mu_{\chi}}{2}& 0
\end{array}
\right).
\label{mmn}
\end{equation}

\hspace{9.1cm}
\rule{-2mm}{0.1mm}\rule{8.7cm}{0.1mm}

\begin{multicols}{2}
\narrowtext

All parameters in Eq.~(\ref{mmn}), but $m^\prime$, are defined in
Eqs.~(\ref{vevs}), (\ref{lambdas}) and (\ref{para}); $g$ and $g^\prime$ denote
the gauge coupling constant of $SU(3)_L$ and $U(1)_N$, respectively. 

The neutralino mass matrix is diagonalized by a $12 \times 12$ rotation unitary 
matrix $N$, satisfying
\begin{equation}
M_{NMD}=N^{*}Y^{0}N^{-1},
\end{equation}
and the mass eigenstates are
\begin{eqnarray}
\tilde{ \chi}^{0}_{i}&=&N_{ij} \Psi^{0}_{j}, \,\ j=1, \cdots ,12.
\label{emasneu}
\end{eqnarray}

We can define the following Majorana spinor to represent the mass eigenstates
\begin{eqnarray}
\Psi(\tilde{ \chi}^{0}_{i})&=& \left( \begin{array}{cc}
             \tilde{ \chi}^{0}_{i} &
	     \bar{ \tilde{\chi}}^{0}_{i} 
\end{array} \right)^T. 
\label{emassneu} 
\end{eqnarray}
As above the subindices $a,b,c$ run over the lepton generations $e,\mu,\tau$.

With the mass matrix in Eq.~(\ref{mmn}), at the tree level we obtain the
eigenvalues (in GeV), 
\begin{equation}
\begin{array}{c}
-4162.22, 3260.48, 3001.11, 585.19,-585.19,453.22,\\-344.14
,283.14,-272.0,
\end{array}
\label{hnm}
\end{equation}
and for the three neutrinos we obtain (in eV)
\begin{equation}
m_1=0,\;m_2\approx -0.01,\; m_3\approx 1.44.
\label{mnus1}
\end{equation}
We have got the values in Eqs.~(\ref{hnm}) and (\ref{mnus1}) by  choosing,
besides the parameters in Eqs.~(\ref{lambdas}) and (\ref{para}), 
$m^\prime=-3780.4159$ GeV. Notice that the coupling constant $g^\prime$ and the parameter $m^\prime$ 
appear only in the mass matrix of the neutralinos, all the other parameters in 
Eq.~(\ref{mmn}) have already been fixed by the charged sector, 
see Eq.~(\ref{clmm}), (\ref{lambdas}) and (\ref{para}). The neutrino masses in
Eq.~(\ref{mnus1}) are of the order of magnitude for LSND and solar neutrino
data. On the other hand if we choose $m^\prime=-3780.4159$ GeV and
$\mu_{0\tau}=2\times10^{-8}$ GeV, we obtain (in eV)  
\begin{equation}
m_1=0.0,\;m_2\approx -5.47\times10^{-5},\; m_3\approx 1.32\times10^{-2},
\label{mnus2}
\end{equation}
which are of the order of magnitude required by the solar and atmospheric
neutrino data. 
Notice also the sensibility of the neutrino masses in $\mu_{0\tau}$ and 
$m^\prime$, and that if $\mu_{0a}=0$, $a=e,\mu,\tau$, all 
neutrinos remain massless. The masses of the charged sector are
insensible to the values of $\mu_{0a}$ for all $a=e,\mu,\tau$ which are suitable
for generating the appropriated different neutrino mass spectra (see the
Appendix~\ref{sec:a2}).  

We have obtained numerically the unitary matrices $E,D$ and $N$ which
diagonalize the mass matrices in Eq.~(\ref{clmm}) and (\ref{mmn}) but we will
not write them explicitly. The charged current is written in the 
mass-eigenstate basis as $\bar{l}_L\gamma^\mu V_{MNS}\nu_LW^-_\mu$ with the
Maki-Nakagawa-Sakata matrix~\cite{mns} defined as 
$V_{MNS}={\cal E}^T_L {\cal N}$, where ${\cal E}$ and ${\cal N}$ are the 
$3\times3$ submatrices of $E$ and $N$, respectively. Hence, we have:
\begin{equation}
V_{MNS}
\approx
\left(\begin{array}{ccc}
1.000 & -0.004 & -0.001 \\
0.001 & -0.001 & 0.003 \\
-0.004 & -0.979 & -0.199
\end{array}
\right).
\label{mns}
\end{equation}

\begin{figure}
\begin{center}
\includegraphics[width=8cm]{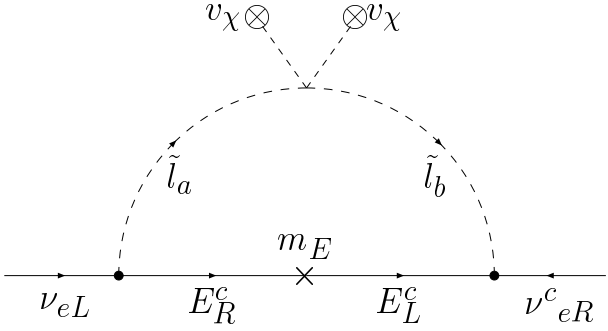}
\end{center}
\caption{
Diagram generating the mass for the lightest neutrino. There is a another
dominant contribution with $v_\chi\to v_{\chi'}$. Each vertex on the left- and
right-side are proportional to $\lambda_{2e\tau}/3$ and $\lambda_{2e\tau}/3 $,
respectively. 
}
\label{fig2}
\end{figure}

Notice that this leptonic mixing matrix is not orthogonal as it must be since
we are omitting the mixture with the heavy charginos and neutralinos and that
one neutrino remains massless at the tree level. We can always rotate the
neutral fields in such a way that the electron neutrino is the one which 
remains massless; or we can also assume $\mu_{0e}=\lambda_{3e}=0$, so that
the electron neutrino decouples from the other neutrinos and neutralinos. In 
this case, diagonal and non-diagonal mass terms in Eq.~(\ref{mmn}) will be 
induced by loop corrections like that in Fig.~\ref{fig2}. Thus, a $3\times3$ 
non-orthogonal mixing matrix will appear in Eq.~(\ref{mns}). Here we 
will only consider the order of magnitude of a mass generated by this process.
 
The massless neutrino can get a mass from the loop correction like that in
Fig.~\ref{fig2} as a consequence of the
Majorana mass term of the neutral lepton in the triplet . This is equivalent to
the mechanism of Ref.~\cite{babuma} but now with a triplet of leptons instead of
a neutral singlet. For instance, the $\lambda_2$ interactions will contribute 
in the left- and right vertices in Fig.~\ref{fig2}:
\end{multicols}
\hspace{-0.5cm}
\rule{8.7cm}{0.1mm}\rule{0.1mm}{2mm}
\widetext

\begin{eqnarray}
{\cal L}^\nu&=&
\frac{\lambda_{2ab}}{3}\left[\overline{\tilde{\eta}^-_{2R}}\,
(\nu_{aL}\tilde{l}_b-
\nu_{bL}\tilde{l}_a)+ \overline{\tilde{\eta}^+_{1R}}\,(\nu_{bL}\tilde{l}^*_a-
\nu_{aL}\tilde{l}^*_b)\right]+\frac{\lambda_{3a}}{3}
\left[\overline{\tilde{\chi}^{++}_R}\,
\nu_{aL}\rho^{++}+ \overline{\tilde{\rho}^{--}_R}\nu_{aL}\chi^{--}-
\overline{\tilde{\chi}^0_R}\nu_{aL}\rho^0-
\overline{\tilde{\rho}^0_R}\nu_{aL}\chi^0\right]\nonumber \\ &+&
\frac{\lambda^\prime_{\alpha ai}}{3}\left[\bar{d}_{\alpha
R}\nu_{aL}\tilde{d}^c_i+ \bar{d}^c_{iR}\nu_{aL}\tilde{d}_\alpha\right]+H.c.,
\label{nusint}
\end{eqnarray}

\hspace{9.1cm}
\rule{-2mm}{0.1mm}\rule{8.7cm}{0.1mm}

\begin{multicols}{2}
\narrowtext
\noindent these interactions generate the lower vertices of Fig.~\ref{fig2}. The
upper vertex are given in Eqs.~(\ref{softnew})--(\ref{ft4}).
With these interactions we can generate the following small mass to the electron
neutrino. In fact, assuming a hierarchy of the masses as in the last subsection, 
we obtain the dominant contribution to the $\nu_e$ mass, up to logarithmic
corrections:
\begin{equation}
m_{\nu_e}\propto\lambda_{2 ea}\lambda_{2 eb}\,E_{ea}E_{eb}\,
V^2_{\tilde{\tau}}\,
\,(v^2_\chi+v^2_{\chi'})
\frac{m_a}{9m^2_{\tilde{\tau}}},
\label{nuemass}
\end{equation}
where all the indices are fixed, $E_{ea}$ is the mixing matrix element defined in
Eqs.~(\ref{m1}) and (\ref{m2}), $V_{\tilde{\tau}}$ denotes the mixing matrix
element in the slepton sector. The charged lepton which gives the
main contribution is the $\tau$ lepton: it has a large mass and the mixing angle
is not too small, in fact $E_{e\tau}=0.004$. Since,
$\lambda_{2e\tau}\lambda_{2e\tau}\sim10^{-6}$ we obtain an electron 
neutrino mass of the order of $10^{-3}$ eV if
$m_{\tilde{\tau}}\approx 4\times10^3 V_{\tilde{\tau}}$ GeV. If
$V_{\tilde{\tau}}=0.02$ we have $m_{\tilde{\tau}}\sim81$ GeV~\cite{pdg}.

\section{Conclusions}
\label{sec:con}

In the nonsupersymmetric 3-3-1 model~\cite{331} with only three scalar
triplets $\eta,\rho\chi$ it is not possible to generate the observed charged
lepton masses. Then, it is necessary to introduce a scalar sextet in order to get
the appropriate masses. When we supersymmetrized the
model and allow $R$-parity breaking interactions we can give to all known
charged leptons and neutrinos the appropriate masses even without the
introduction of a scalar sextet. Of course, in order to cancel anomalies we have
to introduce another set of three triplets $\eta',\rho'\chi'$. In this case
although the correct values for the lepton masses can still be obtained, if the
new VEVs $u^\prime, v^\prime$ and $w^\prime$ are zero, it was shown in
Ref.~\cite{331s} that in order to give mass to all the quarks in the model all
these VEVs have to be different from zero. Hence, we have considered that the
six neutral scalar components got a nonzero VEV. 

As can be seen from Eq.~(\ref{valpara}), the charged lepton masses
arise from a sort of seesaw mechanism since there are small mass parameters, as in
Eq.~(\ref{mus0}), related with $R$-parity breaking interactions, and large ones
as in Eq.~(\ref{para}), related with the mass scale of the supersymmetry
breaking, this can be better appreciated in Eqs.~(\ref{clmmn}). 

The same happens in the neutrino sector, see
Eq.~(\ref{mmnn}). In a supersymmetric version of
the model in which we add the sextet $({\bf6},0)$, there  is a fermionic
non-hermitian triplet under $SU(2)\otimes
U(1)_Y$ that is part of a sextet under $SU(3)\otimes U(1)_N$. This
can also implement a seesaw mechanism for neutrino masses as it was pointed out
in Ref.~(\cite{effop}). The case of a hermitian fermion triplet was considered
in the context of the standard model in Refs.~\cite{rf}. 

It is interesting to note that in the context of MSSM a $Z_2$ 
symmetry~\cite{hall,haber}, 
\begin{equation}
M\to -M,\quad V\to V,\quad X\to X, 
\label{z2}
\end{equation}
where $M,V,X$ is a matter, vector and scalar superfields, respectively,
forbids the $R$-parity breaking terms in Eq.~(\ref{mssmrpv}). 
In the present model it happens the same: the $R$-parity breaking terms in
Eqs.~(\ref{e2}) and (\ref{e22}) are forbidden. Notwithstanding the $Z_3$
symmetry~\cite{haber},
\begin{eqnarray}
\hat{L},\hat{l^c}\to \hat{L},\hat{l^c};\;\; \hat{H_1}\to \hat{H_1},\;
\hat{H_2}\to \hat{H_2};
\nonumber \\
\hat{Q}\to \omega \hat{Q},\; \hat{u^c}\to\omega^{-1}\hat{u^c},
\hat{d^c}\to \omega^{-1}\hat{d^c},
\label{z3}
\end{eqnarray}
where $\omega=e^{2i\pi/3}$, forbids the $B$ violating terms but
allow the $L$ violating ones. This also happens in the present model.   
However, if we introduce an extra discrete $Z^\prime_3$-symmetry, such
that $\hat{L}_e\to-\hat{L}_e$, and all other fields being even under this
transformation, we have that
$\mu_{0e}=\lambda_{2ea}=\lambda_{3e}=\lambda^\prime_{\alpha e i}=0$, at all
orders in perturbation theory. This does not modify the mass matrix in the charged
sector in Eqs.~(\ref{ypm}) and (\ref{clmm}), but forbids  the electron
neutrino to get a mass, at all orders in perturbation theory. 

The present model will induce processes contributing to $\mu\to e\gamma$,
$\tau\to e(\mu)\gamma$, $(g-2)_\mu$, and other exotic decays. However these
processes can be suppressed mainly by the scalar masses since these scalars do
not enter explicitly in the mass matrix at the tree level. Some contributions
to those processes are suppressed by the coupling constants themselves, like
$\lambda_{2}$'s in Eq.~(\ref{lambdas}), other ones which involve
$\lambda_{3\mu},\lambda_{3\tau}$ which are of the order unity can be suppressed
by combining the mixing angles and masses of scalars or charginos sectors.
A more detailed study of this issue will be done elsewhere.
\cite{rf}

In summary, we have analyzed the charged lepton and neutrino masses
in a $R$-parity breaking supersymmetric 3-3-1 model. Unlike the MSSM model 
the electron and its neutrino remain massless at the tree level but gain masses
at the one loop level. The resulting leptonic mixing matrix $V_{MNS}$ is 
non-orthogonal.

\acknowledgments 
This work was supported by Funda\c{c}\~ao de Amparo \`a Pesquisa
do Estado de S\~ao Paulo (FAPESP), Conselho Nacional de 
Ci\^encia e Tecnologia (CNPq) and by Programa de Apoio a
N\'ucleos de Excel\^encia (PRONEX).

\end{multicols}
\hspace{-0.5cm}
\rule{8.7cm}{0.1mm}\rule{0.1mm}{2mm}
\widetext

\appendix

\section{The scalar potential}
\label{sec:a1}

The interactions between the scalars of the theory is given by the scalar
potential that is written as 

\begin{eqnarray}
V_{331}=V_{D}+V_{F}+V_{\mbox{soft}},
\label{ep1}
\end{eqnarray}
where the $V_D$ term is given by
\begin{eqnarray}
V_{D}&=&-{\cal L}_{D}=\frac{1}{2}\left(D^{a}D^{a}+DD\right)\nonumber \\ &=&
\frac{g^{\prime2}}{2}\left(\frac{2}{3} \tilde{Q}^{\dagger}_{1} \tilde{Q}_{1}-
\frac{1}{3} \tilde{Q}^{\dagger}_{\alpha} \tilde{Q}_{\alpha}-
\frac{2}{3} \tilde{u}^{\dagger c}_{i} \tilde{u}^{c}_{1}+
\frac{1}{3} \tilde{d}^{\dagger c}_{i} \tilde{d}^{c}_{1}+
\rho^{\dagger}\rho-\chi^{\dagger}\chi-\rho^{ \prime \dagger}\rho^{ \prime}+
\chi^{ \prime \dagger}\chi^{ \prime} \right)^2 \nonumber \\ 
&+&
\frac{g^2}{8}\sum_{i,j}(\tilde{L}^{\dagger}_{i}\lambda^{a}_{ij} \tilde{L}_{j}+
\tilde{Q}^{\dagger}_{1i}\lambda^{a}_{ij} \tilde{Q}_{1j}+
\eta^{\dagger}_{i}\lambda^{a}_{ij}\eta_{j}
+\rho^{\dagger}_{i}\lambda^{a}_{ij}\rho_{j}
+\chi^{\dagger}_{i}\lambda^a_{ij}\chi_{j} \nonumber \\
&-&
\tilde{Q}^{\dagger}_{\alpha i}\lambda^{* a}_{ij} \tilde{Q}_{\alpha j}
-\eta^{ \prime \dagger}_{i}\lambda^{* a}_{ij}\eta^{ \prime}_{j}
-\rho^{ \prime \dagger}_{i}\lambda^{* a}_{ij}\rho^{ \prime}_{j}
-\chi^{ \prime \dagger}_{i}\lambda^{* a}_{ij}\chi^{ \prime}_{j},
\label{esd}
\end{eqnarray}
the $F$ term is
\begin{eqnarray}
V_F&=&-{\cal L}_F=\sum_{m}F^*_m F_m\nonumber \\ &=&
\sum_{i,j,k}\left[
\left\vert \frac{\mu_{0}}{2} \eta^{\prime}_{i}+ \lambda_{1} \epsilon_{ijk}
\tilde{L}_{j}\tilde{L}_{k}+\frac{2\lambda_{2}}{3} \epsilon_{ijk}
\eta_{j}\tilde{L}_{k}+\frac{\lambda_{3}}{3} \epsilon_{ijk}
\chi_{j}\rho_{k}\right\vert^2 +
\left\vert \frac{\mu_{\eta}}{2} \eta^{\prime}_{i}+\frac{\lambda_{2}}{3} 
\epsilon_{ijk} \tilde{L}_{j}\tilde{L}_{k}+ \frac{f_{1}}{3} \epsilon_{ijk}
\rho_{j}\chi_{k}+\frac{\kappa_{4 \alpha ij}}{3} 
\tilde{Q}_{\alpha}\tilde{d}^{c}_{j}\right\vert^2 \right. \nonumber \\
&+& \left.
\left\vert \frac{\mu_{\rho}}{2}\rho^{\prime}_{i}+\frac{f_{1}}{3}
\epsilon_{ijk}\chi_{j}\eta_{k}+\frac{\kappa_{5 \alpha ij}}{3} 
\tilde{Q}_{\alpha}\tilde{u}^{c}_{j}+\frac{\lambda_{3}}{3} \epsilon_{ijk}
\tilde{L}_{j}\chi_{k}\right\vert^2 +
\left\vert \frac{\mu_{\chi}}{2}\chi^{\prime}_{i}+\frac{f_{1}}{3}\epsilon_{ijk}
\rho_{j}\eta_{k}+\frac{\kappa_{6 \alpha i \beta}}{3} 
\tilde{Q}_{\alpha}\tilde{j}^{c}_{\beta}+\frac{\lambda_{3}}{3} \epsilon_{ijk}
\tilde{L}_{j}\rho_{k}\right\vert^2\right.\nonumber \\ & +&\left.
\left\vert \frac{\mu_{\eta}}{2} \eta_{i}+\frac{f^{\prime}_{1}}{3}
\epsilon_{ijk}\rho^{\prime}_{j}\chi^{\prime}_{k}+\frac{\kappa_{1 ij}}{3} 
\tilde{Q}_{1}\tilde{u}^{c}_{j}\right\vert^2 
+\left\vert \frac{\mu_{\rho}}{2}\rho_{i}+\frac{f^{\prime}_{1}}{3}
\epsilon_{ijk}\chi^{\prime}_{j}\eta^{\prime}_{k}+\frac{\kappa_{2 ij}}{3} 
\tilde{Q}_{1}\tilde{d}^{c}_{j}\right\vert^2+
\left\vert \frac{\mu_{\chi}}{2}\chi_{i}+
\frac{f^{\prime}_{1}}{3}\epsilon_{ijk}\rho^{\prime}_{j}\eta^{\prime}_{k}
+\frac{\kappa_{3 i}}{3} \tilde{Q}_{1}\tilde{J}^{c}\right\vert^2\right.
\nonumber \\ 
&+&\left.
\left\vert
\frac{\kappa_{1 ij}}{3} \eta^{\prime}_{i}\tilde{u}^{c}_{j}+
\frac{\kappa_{2 ij}}{3} \rho^{\prime}_{i}\tilde{d}^{c}_{j}+
\frac{\kappa_{3 i}}{3} \chi^{\prime}_{i}\tilde{J}^{c}
\right\vert^2+
\left\vert
\frac{\kappa_{4 \alpha ij}}{3} \eta_{i}\tilde{d}^{c}_{j}+
\frac{\kappa_{5 \alpha ij}}{3} \rho_{i}\tilde{u}^{c}_{j}+
\frac{\kappa_{6 \alpha i \beta}}{3} \chi_{i}\tilde{j}^{c}_{\beta}+
\frac{\lambda^{\prime}_{\alpha ij}}{3} \tilde{L}_{i}\tilde{d}^{c}_{j}
\right\vert^2 
\right.\nonumber \\ &+&\left.
\left\vert
\frac{\kappa_{1 ij}}{3}\tilde{Q}_{1}\eta^{\prime}_{i}+
\frac{\kappa_{5 \alpha ij}}{3}\tilde{Q}_{\alpha}\rho^{\prime}_{i}+
\frac{\lambda^{\prime \prime}_{ijk}}{3}\tilde{d}^{c}_{i}\tilde{d}^{c}_{k}+
\frac{\lambda^{\prime \prime \prime}_{ij \beta}}{3}\tilde{u}^{c}_{i}
\tilde{j}^{c}_{\beta} \right\vert^2
\right.\nonumber \\ &+&\left.
\left\vert
\frac{\kappa_{2 ij}}{3}\tilde{Q}_{1}\rho^{\prime}_{i}+
\frac{\kappa_{4 \alpha ij}}{3}\tilde{Q}_{\alpha}\eta_{i}+
\frac{\lambda^{\prime}_{\alpha ij}}{3}\tilde{Q}_{\alpha}\tilde{L}_{i}+
\frac{2\lambda^{\prime \prime}_{ijk}}{3}\tilde{d}^{c}_{i}\tilde{u}^{c}_{k}+
\frac{\lambda^{\prime \prime \prime \prime}_{j \beta}}{3}\tilde{J}^{c}
\tilde{j}^{c}_{\beta}\right\vert^2\right],
\label{esf}
\end{eqnarray}

Finally, the soft term is (the following soft-terms do not included the exotic 
quarks)

\begin{eqnarray}
V_{soft}&=&-{\cal L}_{soft}\nonumber \\ &=&
\frac{1}{2} \left( m_\lambda \sum_{w=1}^{8}\lambda^w_A\lambda^w_A+
m^\prime\lambda_B\lambda_B+H.c.\right) 
+m_{L}^{2} \tilde{L}^{\dagger} \tilde{L}+m_{Q_{1}}^{2} 
\tilde{Q}^{\dagger}_{1} \tilde{Q}_{1}+\sum_{\alpha =2}^{3}m_{Q_{\alpha}}^{2} 
\tilde{Q}^{\dagger}_{\alpha} \tilde{Q}_{\alpha}
\nonumber \\ &+& \sum_{i=1}^{3}m_{u_{i}}^2 
\tilde{u}^{c \dagger}_{i} \tilde{u}^{c}_{i}+ \sum_{i=1}^{3}m_{d_{i}}^2 
\tilde{d}^{c \dagger}_{i} \tilde{d}^{c}_{i}+
m^2_{ \eta}\eta^{ \dagger}\eta+m^2_{ \rho}\rho^{ \dagger}\rho+
m^2_{ \chi}\chi^{ \dagger}\chi+
m^2_{\eta^{\prime}}\eta^{\prime \dagger}\eta^{\prime}+
m^2_{\rho^{\prime}}\rho^{\prime \dagger}\rho^{\prime}+
m^2_{\chi^{\prime}}\chi^{\prime \dagger}\chi^{\prime} \nonumber \\ 
&+&\hspace{-2mm}
\left[M^2 \sum_{i=1}^{3}\tilde{L}_{i} \eta^{\dagger}_{i}+ 
\varepsilon_{0} \sum_{i=1}^{3} \sum_{j=1}^{3} \sum_{k=1}^{3} \epsilon_{ijk} 
\tilde{L}_{i} \tilde{L}_{j} \tilde{L}_{k}+
\varepsilon_{1} \sum_{i=1}^{3} \sum_{j=1}^{3} \epsilon_{ijk} \tilde{L}_{i} 
\tilde{L}_{j} \eta_{k} + 
\varepsilon_{3} \sum_{i=1}^{3} \sum_{j=1}^{3} \sum_{k=1}^{3} \epsilon_{ijk} 
\tilde{L}_{i} \chi_{j} \rho_{k} \right.\nonumber \\ &+&\left.
k_1\epsilon_{ijk}\rho_i\chi_j\eta_k+
k^{\prime}_1\epsilon_{ijk}\rho^{\prime}_i\chi^{\prime}_j\eta^{\prime}_k+
\sum_{i=1}^{3}\tilde{Q}_{1}( \zeta_{1i}  \eta^{\prime} \tilde{u}^{c}_{i}+
\zeta_{2i}  \rho^{\prime} \tilde{d}^{c}_{i})+ \sum_{\alpha =2}^{3} 
\tilde{Q}_{\alpha} \left( \sum_{i=1}^{3}\omega_{1\alpha i} \eta 
\tilde{d}^{c}_{i} + \omega_{2\alpha i} \rho \tilde{u}^{c}_{i}\right) \right. 
\nonumber \\ &+& \left. 
\sum_{i=1}^{3}\sum_{j=1}^{3}\sum_{k=1}^{3}\varsigma_{1ijk} \tilde{d}^{c}_{i} 
\tilde{d}^{c}_{j} \tilde{u}^{c}_{k}+H.c. \right].
\label{ess}
\end{eqnarray}
\rule{-2mm}{0.1mm}\rule{8.7cm}{0.1mm}

\begin{multicols}{2}
\narrowtext

\section{Numerical analysis of mass matrices}
\label{sec:a2}

Here we show explicitly the numerical values of each entry of the mass matrices
in Eqs.~(\ref{clmm}) and (\ref{mmn}) using the parameters given in
Eq.~(\ref{valpara}) and $m^\prime=-3780.4159$ GeV.
For the charged sector we have 
\end{multicols}
\hspace{-0.5cm}
\rule{8.7cm}{0.1mm}\rule{0.1mm}{2mm}
\widetext

{\scriptsize
\begin{equation}
X= \left( \begin{array}{ccccccccc} 
0.0&-0.005&-0.005&0.0&0.0&0.0&0.0&-0.024&0.0 \\ 
0.005&0.0&-1.851&0.0&0.0&0.0&0.0&-235.702&0.0 \\
0.005&1.851&0.0&0.0&0.0&0.0&0.0& -235.702&0.0 \\
0.0&0.0&0.0&3000.0&0.0&-0.462&0.0&80.071&0.0 \\
0.0&0.0&0.0&0.0&3000.0&0.0&9.237&0.0&-923.707 \\
0.0&0.0&0.0&9.237&0.0& -150.0&0.0&59.868&0.0 \\
0.0&0.0&0.0&0.0&-0.462&0.0&-150.0&0.0&-0.236 \\
0.0&0.0&0.0&-0.462&0.0&471.405&0.0&-250.0&0.0 \\ 
-0.004&-40.864&-40.864&0.0&461.854&0.0&-10.379&0.0&-350.000
\end{array}
\right).
\label{clmmn}
\end{equation}
}
\rule{-2mm}{0.1mm}\rule{8.7cm}{0.1mm}
\begin{multicols}{2}
\narrowtext
and for the neutral sector: 
\end{multicols}
\hspace{-0.5cm}
\rule{8.7cm}{0.1mm}\rule{0.1mm}{2mm}
\widetext
{\scriptsize
\begin{equation}
\hspace{-13 mm}
Y^{0}= \left( \begin{array}{cccccccccccc}
0.0&0.0&0.0&0.0&0.0&0.0&0.0&0.0&0.024&0.0&0.004&0.0 \\
0.0&0.0&0.0&0.0&0.0&0.0&0.0&0.0&235.702&0.0&40.864&0.0 \\
0.0&0.0&0.0&0.0&0.0&0.0&0.0&0.0&235.702&0.0&40.864&0.0 \\
0.0&0.0&0.0&3000.0&0.0&0.0&6.532&-0.327&-56.619&0.327&0.0&0.0 \\
0.0&0.0&0.0&0.0&3000.0&0.0&3.771&-0.189&32.689&-0.189&-377.102&754.204 \\
0.0&0.0&0.0&0.0&0.0&-3780.416&0.0&0.0&99.391&-0.573&-573.290&1146.580 \\
0.0&0.0&0.0&6.532&3.771&0.0&0.0&-150.0&-59.868&0.0&10.379&0.0 \\
0.0&0.0&0.0&-0.327&-0.189&0.0&-150.0&0.0&0.0&-471.405&0.0&0.236 \\
0.024&235.702&235.702&-56.619&32.689&99.391&-59.868&0.0&0.0&-250.0&-1.197&0.0 \\
0.0&0.0&0.0&0.327&-0.189&-0.573&0.0&-471.405&-250.0&0.0&0.0&-0.236 \\
0.004&40.864&40.864&0.0&-377.102&-573.290&10.379&0.0&-1.197&0.0&0.0&-350.0 \\
0.0&0.0&0.0&0.0&754.204&1146.580&0.0&0.236&0.0&-0.236&-350.0&0.0 \\
\end{array}
\right).
\label{mmnn}
\end{equation}
}
\rule{-2mm}{0.1mm}\rule{8.7cm}{0.1mm}

\begin{multicols}{2}
\narrowtext

Here we show that the relevant parameters for the leptons masses are
$\lambda_{2,3}$. We note that there are four type of parameters in  the mass
matrices in Eqs.~(\ref{valpara}). Firstly we have the dimensionless Yukawa
couplings in the usual leptons, $\lambda_{2,3}$ and in the supersymmetric
partners,  $f_1,f^\prime_1$. We also have the mass dimension parameters
$\mu_{0a}$ and $\mu_{\eta,\rho,\chi}$ and $m_\lambda$ and $m^\prime$ which are
soft terms in Eq.~(\ref{ess}). Of all these parameters we expect that the
relevant ones in the charged lepton and neutrinos are $\lambda_{2,3}$. To show
this we consider several choices of the parameters as follows:
(Below all masses are in GeV)\\
{\bf case 1:}
\begin{eqnarray}
\lambda_{2e\mu}=0.0,\;\lambda_{2e\tau}=0.0,\;\lambda_{2\mu\tau}=0.0,
\nonumber \\
\lambda_{3e}=0.0,\; \lambda_{3\mu}=0.0,\;\lambda_{3\tau}=0.0,\nonumber \\
\mu_{0e}=\mu_{0\mu}=0.0;\mu_{0\tau}=0.0, \,\ \mbox{(in GeV)}.
\label{c1}
\end{eqnarray}
Charged sector masses:\\
3186.03, 3001.10, 557.17, 196.55,149.30, 16.85, 0, 0, 0. \\
Neutral sector masses:\\
-4162.22, 3260.47, 3001.10, 557.79, -557.17, 450.14, -330.68, 17.18, -17.01, 
0, 0, 0.

{\bf case 2:} 
\begin{eqnarray}
\lambda_{2e\mu}=0.0,\;\lambda_{2e\tau}=0.0,\;\lambda_{2\mu\tau}=0.0,
\nonumber \\
\lambda_{3e}=0.0,\; \lambda_{3\mu}=0.0,\;\lambda_{3\tau}=0.0,\nonumber \\
\mu_{0e}=\mu_{0\mu}=0.0;\mu_{0\tau}=2 \times 10^{-8}, \,\ \mbox{(in GeV)}
\label{c2}
\end{eqnarray}
Charged sector masses:\\
3186.03,3001.10,557.17,196.55,149.30,16.85,1.92$\times 10^{-12}$, 0, 0.\\ 
Neutral sector masses:\\
-4162.22, 3260.47, 3001.10, 557.79, -557.17, 450.14, -330.68, 17.18, -17.01, 
2.80 $\times 10^{-21}$, 0, 0.
 
{\bf case 3:} 
\begin{eqnarray}
\lambda_{2e\mu}=0.0,\;\lambda_{2e\tau}=0.0,\;\lambda_{2\mu\tau}=0.0,
\nonumber \\
\lambda_{3e}=0.0001,\; \lambda_{3\mu}=1.0,\;\lambda_{3\tau}=1.0,\nonumber \\
\mu_{0e}=\mu_{0\mu}=0.0;\mu_{0\tau}=0.0, \,\ \mbox{(in GeV)}.
\label{c3}
\end{eqnarray}
Charged sector masses:\\
3186.05, 3001.11, 584.85, 282.30, 149.41, 204.55, 2.10$\times 10^{-10}$, 0,
0.\\
Neutral sector masses:\\
-4162.22, 3260.47, 3001.10, 585.18, -585.18, 453.22, -344.14, 283.14,
 -271.99,  
1.23 $\times 10^{-11}$, 0, 0.
 
{\bf case 4:} 
\begin{eqnarray}
\lambda_{2e\mu}=0.001,\;\lambda_{2e\tau}=0.001,\;\lambda_{2\mu\tau}=0.393,
\nonumber \\
\lambda_{3e}=0.0,\; \lambda_{3\mu}=0.0,\;\lambda_{3\tau}=0.0,\nonumber \\
\mu_{0e}=\mu_{0\mu}=0.0;\mu_{0\tau}=0.0, \,\ \mbox{(in GeV)}.
\label{c4}
\end{eqnarray}
Charged sector masses:\\
3186.03, 3001.10, 557.17, 196.55,149.30, 16.85, 1.85, 1.85, 0.\\
Neutral sector masses:\\
-4162.22, 3260.47, 3001.10, 557.79, -557.17, 450.14, -330.68, 17.18, -17.01, 
0, 0, 0.

{\bf case 5:} 
\begin{eqnarray}
\lambda_{2e\mu}=0.001,\;\lambda_{2e\tau}=0.001,\;\lambda_{2\mu\tau}=0.393,
\nonumber \\
\lambda_{3e}=0.0001,\; \lambda_{3\mu}=1.0,\;\lambda_{3\tau}=1.0,\nonumber \\
\mu_{0e}=\mu_{0\mu}=0.0;\mu_{0\tau}=0.0, \,\ \mbox{(in GeV)}
\label{c5}
\end{eqnarray}
Charged sector masses:\\
3186.03, 3001.11, 584.85, 282.30, 204.55, 149.41, 1.78, 0.105, 0.\\
Neutral sector masses:\\
-4162.22, 3260.47, 3001.10, 585.19, -585.19, 453.22, -344.14, 283.14, -271.99, 
1.23 $\times 10^{-11}$, 0, 0.

Notice that the values of the masses in the charged sector are not significantly
affected by the values of $\mu_{0a}$.

\end{multicols}

\newpage

\end{document}